\documentclass{moriond}
 \usepackage{amsbsy}
 \usepackage{amssymb}
 \usepackage{amsmath}
 \usepackage{ wasysym }
 
\def\Journal#1#2#3#4{{#1} {\bf #2}, #3 (#4)}

\def\NIMA{{\em Nucl. Instrum. Methods} A}

\def\PRL{\em Phys. Rev. Lett.}
\def\PRD{{\em Phys. Rev.} D}
\def\PRC{{\em Phys. Rev.} C}

\def\JINST{{\em JINST}}
\def\JOPG{{\em J. Phys.} G{\em : Nucl. Part. Phys.}}

\def\be{\begin{equation}}
\def\ee{\end{equation}}
\def\bea{\begin{eqnarray}}
\def\eea{\end{eqnarray}}

\newcommand{\STEREO}{\textsc{Stereo}}
\newcommand{\U}{$^{235}$U}

\newcommand{\Photo}{}

\begin{document}
\vspace*{4cm}
\title{RESULTS OF STEREO AND PROSPECT, AND STATUS OF STERILE NEUTRINO SEARCHES}

\author{ M. LICCIARDI, on behalf of the \STEREO~collaboration }

\address{Univ. Grenoble Alpes, CNRS, Grenoble INP, LPSC-IN2P3, 38000 Grenoble, France}

\maketitle\abstracts{Reactor neutrinos have been an intense field of investigation for the last decade. Two anomalies are discussed in this document. First, a status of the sterile neutrino searches by \STEREO, PROSPECT and DANSS is presented. The best-fit parameters of active-to-sterile oscillations from the Reactor Antineutrino Anomaly are strongly rejected by these experiments. Second, the analyses of the shape anomaly ("5 MeV bump") by \STEREO~and PROSPECT, both using a virtually pure~\U~neutrino flux, are detailed. Results show a significant excess of events at 5-6~MeV and indicate that the bump observed at commercial reactors is not specific to a particular isotope but rather shared among U and Pu.}

\section{Introduction}

\hspace{0.5cm} The Reactor Antineutrino Anomaly (RAA) \cite{RAA} appeared in 2011, following a revision of the predicted neutrino fluxes for the main isotopes in nuclear fuel ($^{235}$U, $^{238}$U, $^{239}$Pu, $^{241}$Pu) \cite{Mueller,Huber}. The upward reevaluation of predicted fluxes resulted in a data-to-prediction deficit of about 5\%. One hypothesis to explain this deficit consists in oscillations to a sterile neutrino state, since sterile neutrinos are not observable in detectors. The survival probability writes for $L\lesssim 1$ km: \begin{equation}\label{eqn:proba}
P_{ee} = 1 - \cos^4 (\theta_{14}) \sin^2 (2\theta_{13}) \sin^2 \left(\frac{\Delta m^2_{31} L}{4E} \right) - \sin^2 (2\theta_{14}) \sin^2 \left(\frac{\Delta m^2_{14} L}{4E} \right)
\end{equation}
with $E$ the neutrino energy, $L$ the distance from source to detector, $\theta_{13}$ and $\Delta m^2_{13}$ the mixing angle and mass splitting in the 3-neutrino paradigm, $\theta_\mathrm{14}$ and $\Delta m^2_\mathrm{14}$ the parameters ruling the oscillation to the sterile state. Based on data available at the time, the RAA best-fit parameters are of the order of $\sin^2 2\theta_\mathrm{new} \sim 0.1$ and $\Delta m^2_\mathrm{new} \sim 2~ \mathrm{eV}^2$. Notably, this mass splitting translates into an oscillation length from 2 to 10 meters (depending on the neutrino energy), which is the typical length on which oscillations would develop. Therefore, a new generation of neutrino detectors were designed to study such oscillations, combining two key requirements: (i) a distance from detector to reactor core of about 10 m, to be able to probe the RAA hypothesis; (ii) a segmented detector covering several baselines over $\sim 2$ m, so that oscillations could develop inside the detector and be seen by comparing spectra from the detector's subparts. Experiments operating such detectors are, for instance, \STEREO, PROSPECT and DANSS.

Another intriguing aspect of the reactor antineutrino spectrum is the excess of events observed by experiments running at commercial reactors (Daya Bay \cite{DB}, RENO \cite{RENO}, Double Chooz \cite{DC}) around 5 MeV, so-called "5 MeV bump". In order to disentangle the contribution from the several isotopes present in low-enriched nuclear fuel, pure \U~measurements from research reactors are helpful. \STEREO~and PROSPECT have thus performed analyses of the spectrum shape, and have formed a joint collaboration to increase the sensitivity of their respective analyses.

\section{Experimental context}
\subsection{\STEREO, {\em PROSPECT} and {\em DANSS} experiments}
\hspace{0.5cm} Three experiments will be presented in this document: i) \STEREO~\cite{ST}, operated near the High Flux Reactor at Institut Laue-Langevin in Grenoble, France; ii) PROSPECT \cite{PR}, operated near the High Flux Isotope Reactor at the Oak Ridge National Laboratory, USA; iii) DANSS~\cite{DANSS}, operated below a commercial reactor core at the Kalininskaya nuclear power plant in Russia. Tab. \ref{tab:exp} compares some important instrumental characteristics of these three experiments.

\begin{table}[!h]
\caption[]{Comparison of key instrumental aspects of the \STEREO, PROSPECT and DANSS experiments.}
\label{tab:exp}
\vspace{0.4cm}
\begin{center}
\begin{tabular}{|c|c|c|c|}
\hline
& \STEREO & PROSPECT & DANSS \\ \hline
Reactor power & 58 MW$_\mathrm{th}$ & 85 MW$_\mathrm{th}$ & 3 GW$_\mathrm{th}$\\
\U~enrichement & 93 \% (HEU) & 93 \% (HEU) & Low (LEU) \\
$\overline{\nu}_e$ spectrum & $>99\%$ from \U & $>99\%$ from \U & Mixed isotopes \\
Core size & $h=80$cm, $\diameter=40$cm & $h=50$cm, $\diameter=45$cm & $h=3.5$m, $\diameter=3.1$m \\ \hline
Detector structure & Segmented (6 cells) & Segmented & On a movable platform \\
Baselines & 9.4 to 11.2 m & 6.7 to 9.3 m & 10.9, 11.9 and 12.9 m \\ \hline
Detection principle & Gd-doped liq. scint. & $^6$Li-doped liq. scint. & Gd-coated plastic scint. \\ 
$\overline{\nu}_e$ rate & 360 / day & 500 / day & 5 000 / day \\
S/B &  $\simeq 0.8$ & $\simeq 1.4$ & $\sim 20$ \\ \hline
\end{tabular}
\end{center}
\end{table}

\subsection{Analysis key points: \STEREO~as an illustration}
\hspace{0.5cm} Let us now illustrate some key aspects of the analysis using \STEREO~as an example. Reactor antineutrinos are detected through the inverse beta decay (IBD) reaction: $\overline{\nu}_e + p \rightarrow e^+ + n$. This produces two coincident signals in the detector: i) a prompt signal from the $e^+$ kinetic energy and annihilation, and ii) a delayed signal from the neutron capture on a Gd nucleus.

The prompt signal is related to the neutrino energy as $E_\mathrm{pr} = E_\nu - 0.782$ MeV. An accurate energy reconstruction is crucial for the oscillation analysis since they develop as a function of $L/E$, and for the spectral analysis as well. Most importantly the reconstructed energy has to be well reproduced in simulation: the quantity of interest is the \textit{energy scale}, namely the data/MC ratio of reconstructed energies $E_\mathrm{rec}^\mathrm{data} / E_\mathrm{rec}^\mathrm{MC}$. The control of the energy scale is done in 3 steps. First, a set of calibration sources are deployed in each cell, at 5 different heights, and cover energies from 0.5 MeV to $\simeq$8 MeV. Second, the $\beta$-decay of cosmic-muon-induced $^{12}$B provides a continuous spectrum up to $Q_\beta = 13.4$ MeV. Finally, a joint fit of the data/MC residuals from sources and boron constrains the ratio $E_\mathrm{rec}^\mathrm{data} / E_\mathrm{rec}^\mathrm{MC}$ to be in agreement with 1 within $\pm 1\%$ accuracy across all energies.

The delayed signal originates from the de-excitation cascade of the excited Gd nucleus after $n$ capture. The computation of the capture efficiency is of prime importance for the absolute measurement of the IBD rate which provides a measurement of the deficit with respect to the reference Huber-Mueller (HM) model \cite{Huber,Mueller}. It is done with an AmBe source circulated in and around the detector, systematic uncertainties being evaluated with data/MC comparison. For this, the FIFRELIN software provides improved modelling of the de-excitation cascades \cite{FIFRELIN}, the result of which is made available to the community. In addition, correction factors acount for spatial data/MC discrepancies that could be related to neutron physics at the edges of the detector, or Gd non-uniformity. This careful evaluation allows \STEREO~to measure the rate deficit of the $^{235}$U spectrum as \begin{equation}
( 5.2 \pm 0.8 [\mathrm{stat}] \pm 2.3 [\mathrm{syst}] \pm 2.3 [\mathrm{model}] ) \%
\end{equation} which is the most accurate result for HEU experiments \cite{RatePaper}.

A final key point in the analysis comes from the reactor's alternation of reactor-on and reactor-off periods. Cosmogenic background is measured in reactor-off periods and used as a model for such a background in the reactor-on collected data. Pulse Shape Discrimination then allows to keep a signal-to-background ratio of order 1, leading to the sterile neutrino search and the spectral analysis presented in Section~\ref{scn:OA} and \ref{scn:shape}. Note that the following results use only the first 2 phases of \STEREO's collected data, with a doubling of statistics yet to come.

\section{Sterile neutrino searches}\label{scn:OA}

\hspace{0.5cm} \STEREO, PROSPECT and DANSS use different analysis methods, but all rely on the same principle: looking for oscillations by comparing spectra at different baselines (and not comparing the data to a model). Analyses can then be model-independent and provide more reliable results.

In order to do so, \STEREO~introduces free $\phi_j$ parameters in its $\chi^2$ minimization \cite{LongPaper}:\begin{equation}
\chi^2 = \sum_{l=1}^{N_\mathrm{cells}}\sum_{j=1}^{N_\mathrm{Ebins}} \left(\frac{A_{l,j} - \phi_j M_{l,j}}{\sigma_j}\right)^2 + \mathrm{pull~terms}
\end{equation} with $A_{l,j}$ the measured neutrino rate in cell $l$ and energy bin $j$, and $M_{l,j} = M_{l,j}(\sin \theta_{14}, \Delta m_{14}^2; \boldsymbol{\alpha})$ the predicted rate as a function of the active-to-sterile oscillation parameters from Eq.~\ref{eqn:proba} and nuisance parameters $\boldsymbol \alpha$. Any bias in the initial model in energy bin $j$ would be absorbed by the $\phi_j$ in this bin: then, only cell-to-cell differences in the oscillated prediction $M_{l,j}$ remain relevant.

A similar method is pursued by PROSPECT \cite{PRosc}. Specifically, the $\phi_j$ are not free parameters anymore but are set to \begin{equation}
\phi_j = \sum_{l=1}^{N_\mathrm{baselines}} A_{l,j} \, / \sum_{l=1}^{N_\mathrm{baselines}} M_{l,j}
\end{equation} using the ratio of the baseline-summed measured and predicted spectra.

Finally, the analysis of the DANSS collaboration uses the ratio $R_j$ of spectra between bottom ($L=12.9$ m) and top ($L=10.9$ m) positions \cite{DANSSanalysis}. The $\chi^2$ then writes \begin{equation}
\chi^2 = \sum_{j=1}^{N_\mathrm{Ebins}} \left( \frac{R^\mathrm{obs}_j - k\times R^\mathrm{pred}_j}{\sigma_j} \right)^2
\end{equation} with $R^\mathrm{obs}_j$ and $R^\mathrm{pred}_j$ the observed and predicted ratios, and $k$ a normalization parameter to consider only differences in positron energy shapes.\\

\begin{figure}[!ht]
\centering
\begin{minipage}{0.46\linewidth}
\centerline{\includegraphics[height=0.95\linewidth]{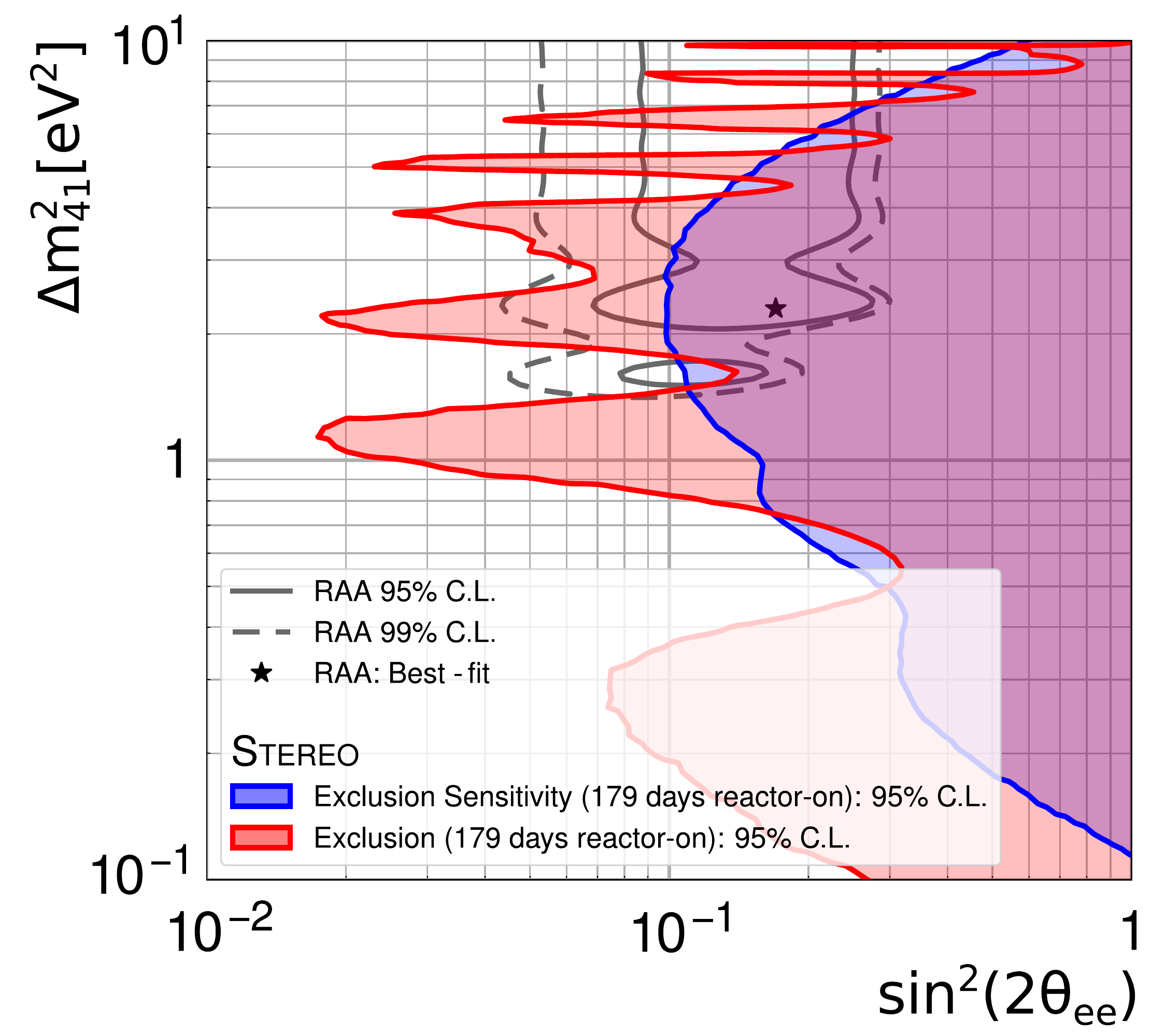}}
\end{minipage}
\hfill
\begin{minipage}{0.46\linewidth}
\centerline{\includegraphics[height=0.95\linewidth]{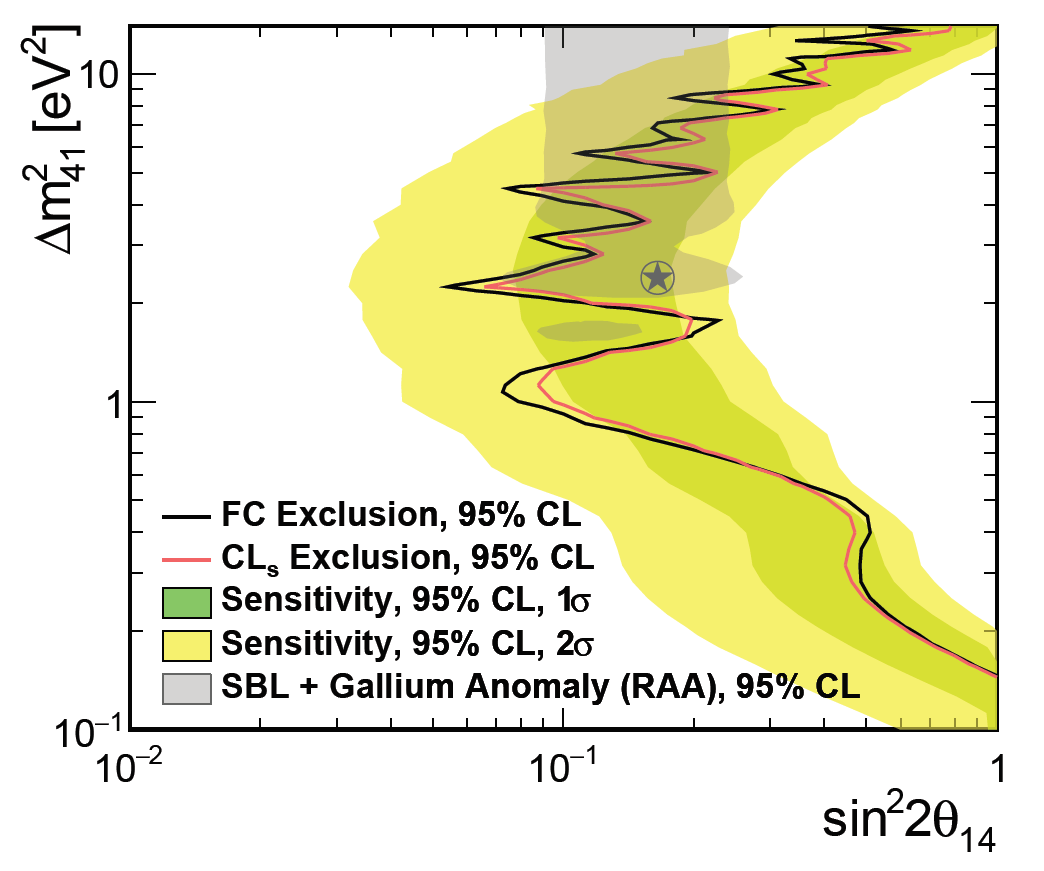}}
\end{minipage}

\begin{minipage}{0.5\linewidth}
\centerline{\includegraphics[height=0.95\linewidth]{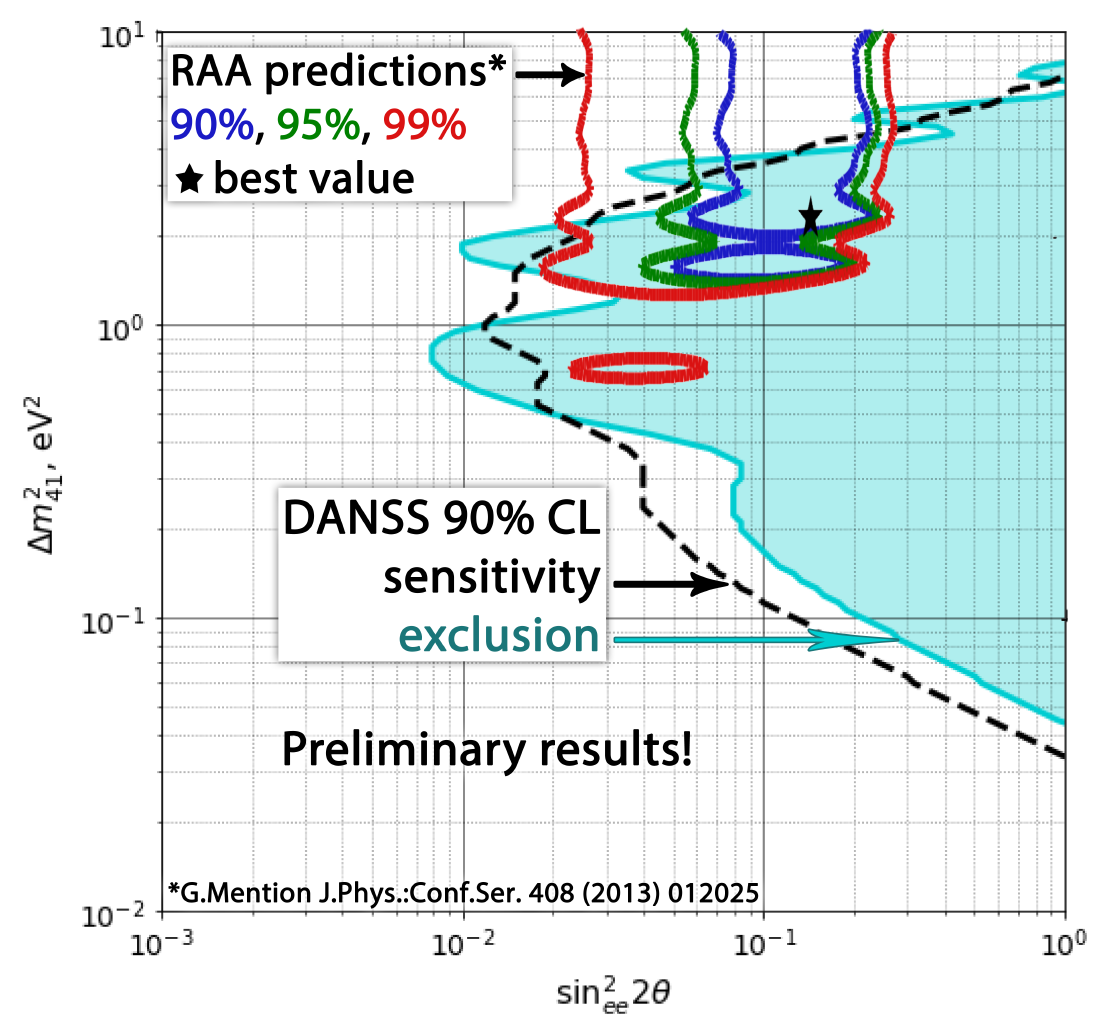}}
\end{minipage}
\caption[]{Exclusion and sensitivity contours from (top left) \STEREO~\cite{LongPaper}, (top right) PROSPECT \cite{PRosc} and (bottom) DANSS \cite{DANSSosc}, showing strong rejection of the RAA best-fit point as well as a large surrounding region.}
\label{fig:osc}
\end{figure}

The exclusion contours from the 3 experiments are gathered in Figure \ref{fig:osc}, using either the CLs (PROSPECT, DANSS) or the usual $\Delta \chi^2$ method (\STEREO, PROSPECT). Importantly for the latter, since $\Delta \chi^2$ does not follow a standard $\chi^2$ law with 2 d.o.f., the Feldman-Cousins procedure (buiding $\Delta \chi^2$ p.d.f.s via MC toys) is used. The best-fit point of the RAA is strongly excluded: at $> 99\%$ C.L. by \STEREO, at $>95\%$ C.L. by PROSPECT, and at $> 5\sigma$ by DANSS. A large portion of the RAA 95\% C.L. region is excluded as well. The only remaining region of the parameter space still unrejected corresponds to $\Delta m^2 \gtrsim 5~\mathrm{eV}^2$ where experiments at $O(10~\mathrm{m})$ have little sensitivity. 

In this region of high-frequency oscillations, only an averaged effect could be observable in detectors, and no cell-to-cell distortions are to be expected. Oscillations to a sterile state would only appear as a deficit to a certain model, and are thus beyond the model-independent searches discussed here. One would need, for instance, to add a model-dependent constraint on the normalization.

\section{Spectral analysis}\label{scn:shape}
\hspace{0.5cm} For the experiments at HEU reactors, the spectral analysis has been developped for two main purposes: i) to investigate the \U~contribution to the event excess around 5 MeV, and ii) to provide an experimental reference to the community, since the current prediction fails to reproduce the experimental data in this specific region. Therefore, the measured energy spectra have been unfolded and expressed as a function of the antineutrino energy.

\subsection{Analysis of \STEREO~data}
\hspace{0.5cm} The unknown unfolded spectrum is parametrized by a set of parameters $\Phi_i$ (one per $E_\nu$ bin $i$) and is folded into measured-energy space through the experimental response matrix $R$. There, it is fitted against the measured spectrum $D$, using one of the following two methods: \begin{equation}\label{eqn:ST-1}
\chi^2 (\Phi) = \big(R \Phi - D\big)^T V_D^{-1} \big(R  \Phi - D\big)  + r \cdot \mathcal{R}_1(\Phi)
\end{equation}
or
\begin{equation}
 \chi^2(\Phi) = \big(R(\vec{\alpha}) \Phi - D\big)^T V_\mathrm{stat}^{-1} \big(R(\vec{\alpha})  \Phi - D\big)+|\vec{\alpha}|^2  + r \cdot \mathcal{R}_1(\Phi).
\end{equation}

The first method uses the total covariance matrix $V_D$ in the $\chi^2$ computation, while systematic uncertainties are implemented using nuisance parameters in the second. As statistical noise is enhanced by unfolding processes, both uses a Tikhonov-like regularization term \begin{equation}
\mathcal{R}_1(\Phi) = \sum_i \left(  \frac{\Phi_{i+1}}{\Phi_{i+1}^0} -  \frac{\Phi_i}{\Phi_i^0} \right)^2
\end{equation} to smooth the first derivative of the deviation between the unfolded spectrum and a prior shape $\Phi^0$ (set to the shape of the HM model \cite{Huber,Mueller}). The strength of this smoothing can be varied through the tunable parameter $r$. In the first method, $r$ is set following the General Cross-Validation prescription~\cite{GCV}; in the second method, $r$ is empirically chosen to ensure that the choice of the prior shape has negligible impact on the unfolded spectrum \cite{ShapePaper}. The compatibility between the two methods has been validated on a pseudo-data set, and excellent agreement was found. Finally, biases have been studied on various toy models (including models with a bump) and show that the unfolding is well controlled, as biases remain $<1\%$ across all the energy range. 

\begin{figure} [!ht]
\centering
\centerline{\includegraphics[width=0.65\linewidth, trim={0 0.5cm 0 0},clip]{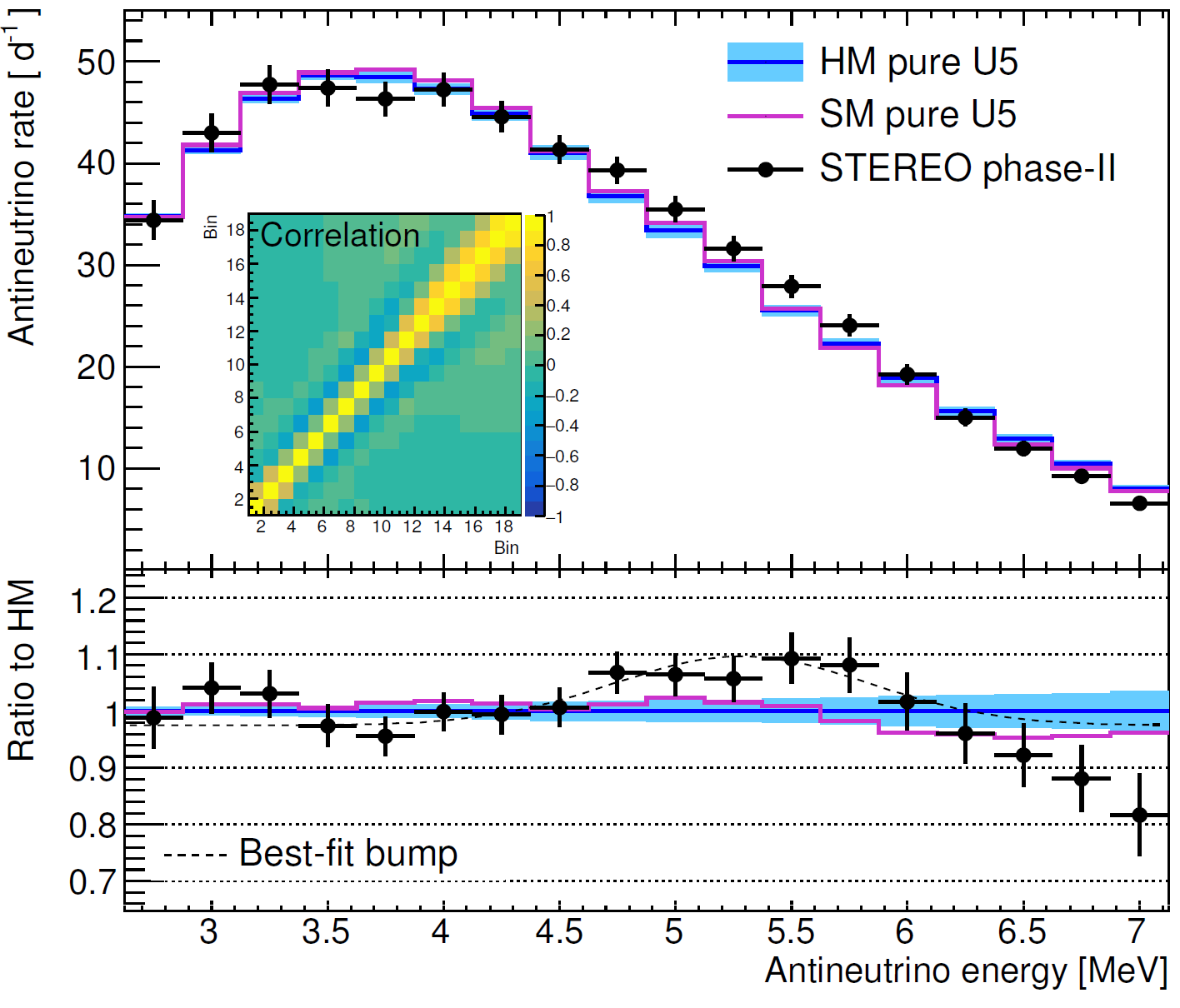}}
\caption[]{(Top) Unfolded \U~spectrum along with area-normalized Huber-Mueller (HM) and Summation Model (SM) \cite{SM} predictions. The non-trivial correlation matrix is displayed. (Bottom) Ratios to HM prediction. Adapted from Ref. \cite{ShapePaper}.}
\label{fig:ST-spectrum}
\end{figure}

The unfolded spectrum is shown in Fig.~\ref{fig:ST-spectrum}. The agreement with the area-normalized HM model is $\chi^2 / \mathrm{ndf} = 26.7/18$ (p-value: 0.084). When fitted by a gaussian, the excess of events around 5-5.5 MeV gives an amplitude of $(12.1 \pm 3.4) \%$. The significance of the bump observation in a \U~spectrum by \STEREO~reaches thus 3.5$\sigma$. However, no preference can be inferred about the isotopic origin of the bump observed by Daya Bay \cite{DB} with commercial reactors. A bump shared among isotopes (resp. a pure-\U~bump) would correspond to an amplitude of $\sim 9\%$ (resp. $\sim 16\%$) in \STEREO~data.

\subsection{Analysis of {\em PROSPECT} data}
\hspace{0.5cm} The investigation for the bump is done on the measured-energy spectrum by PROSPECT. The analysis focuses on the isotopic origin of the bump observed by Daya Bay \cite{DB}. A gaussian with mean 5.678~MeV and width 0.562~MeV (corresponding to Daya Bay's observation) is folded through the response matrix, and its amplitude is fitted againt PROSPECT data \cite{PRosc}.

\begin{figure} [!ht]
\centering
\centerline{\includegraphics[width=0.62\linewidth]{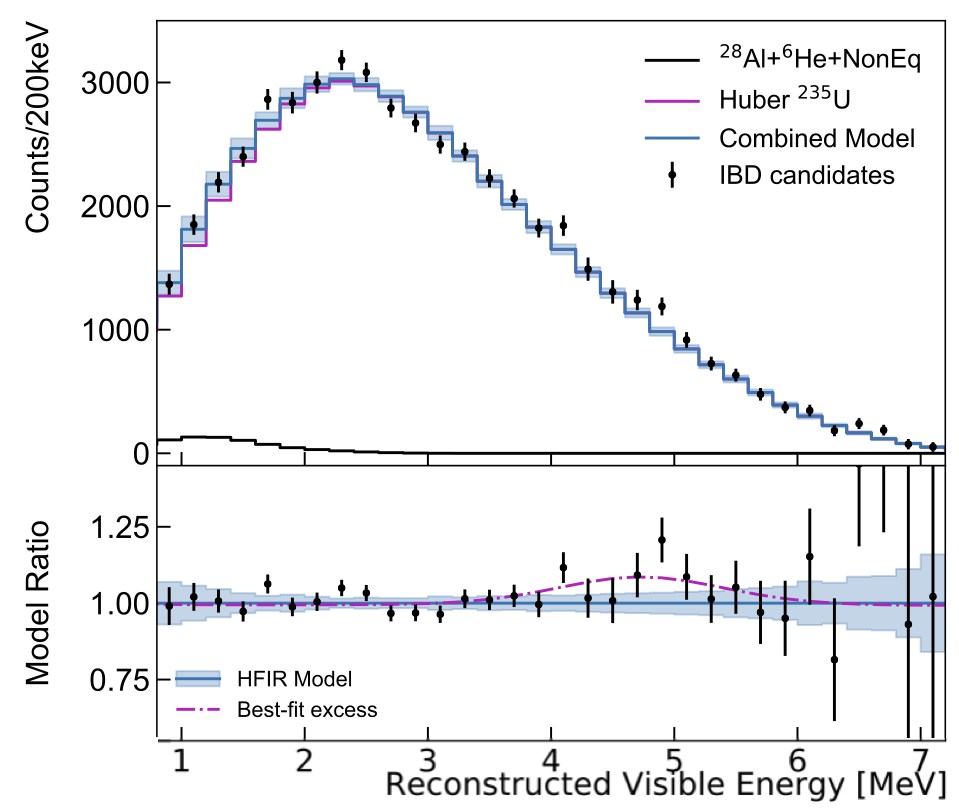}}
\caption[]{ Measured-energy spectrum observed by the PROSPECT experiment \cite{PRosc}. The event excess is fitted by a gaussian with mean and width set to correspond to Daya Bay's observation.}
\label{fig:PR-spectrum}
\end{figure}

PROSPECT spectrum and the best-fit excess are displayed on Fig.~\ref{fig:PR-spectrum}. The  scale factor, ratio of PROSPECT's to Daya Bay's amplitudes, has a best-fit value of $0.84 \pm 0.39$. It is consistent with 1, indicating that \U~does not play a specific role on the bump origin with respect to the other isotopes present in LEU neutrino spectra. The no-\U~bump and all-\U~bump hypotheses, corresponding to scale factors of 0 and 1.78 respectively, are both disfavoured at more than $2\sigma$.

\subsection{Joint analysis by \STEREO~and {\em PROSPECT}}
\hspace{0.5cm} As both \STEREO~and PROSPECT spectral analyses are statistically limited, both collaborations have worked together on a joint analysis to increase the physics reach of the experiments. \STEREO's unfolding methods have been extended to perform the joint unfolding of the data sets; for instance, the first method now writes
\begin{align} 
\chi^2(\Phi)& = \big(R_\mathrm{ST} \Phi - D_\mathrm{ST}\big)^T V_\mathrm{ST}^{-1} \big(R_\mathrm{ST} \Phi - D_\mathrm{ST}\big) \nonumber \\ 
&+ \big(\alpha \cdot R_\mathrm{PR} \Phi - D_\mathrm{PR}\big)^T V_\mathrm{PR}^{-1} \big( \alpha \cdot R_\mathrm{PR} \Phi - D_\mathrm{PR}\big)  + r \cdot \mathcal{R}_1(\Phi) 
\label{eqn:joint-fit}
\end{align}
where PR and ST are used to refer to each experiment, and $\alpha$ is a free-floating scaling parameter accounting for the different normalizations of the experiments. 

Another unfolding framework has been developped within the PROSPECT collaboration, based on the Wiener-SVD method \cite{WienerSVD}. The unfolding is done by using the Singular Value Decomposition to invert the response matrix. A filter, optimized according to the signal-to-background ratio, is applied in this inversion process to mitigate the noise amplification due to small diagonal elements in the SVD decomposition.

\begin{figure} [!ht]
\centering
\centerline{\includegraphics[width=0.55\linewidth]{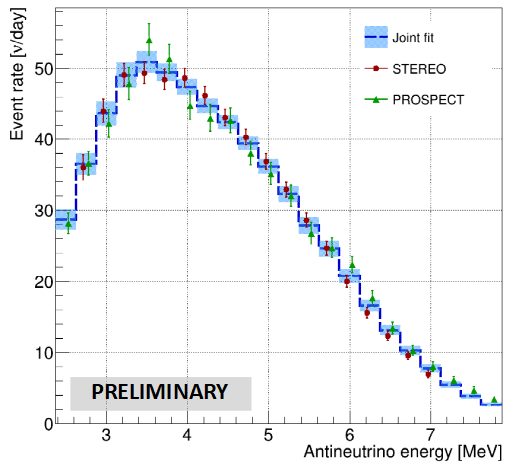}}
\caption[]{Joint fit of \STEREO~and PROSPECT data, along with the unfolded spectra of each experiments (all unfolding performed with the method of Eqn.~\ref{eqn:joint-fit}).}
\label{fig:joint-fit}
\end{figure}

\STEREO's and PROSPECT's unfolding methods have been compared on the single unfolding of \STEREO~or PROSPECT data, and show excellent agreement. Moreover, a $\chi^2$ test between the spectra from the two experiments (unfolded with the method of Eqn. \ref{eqn:joint-fit}) gives $\chi^2 / \mathrm{ndf} = 22.3 / 17$ (p-value: 0.17). Since the two experiments are statistically compatible, it is fully relevant to attempt a joint fit combining both of them. A preliminary $E_\nu$ spectrum from the joint unfolding of \STEREO~and PROSPECT data is shown in Fig.~\ref{fig:joint-fit}. As suggested in Ref. \cite{WienerSVD}, a \textit{filter matrix} encoding all unfolding biases will be provided with the unfolded spectrum in the forthcoming publication. 

This \U~spectrum extracted by \STEREO~and PROSPECT, from HEU reactor cores, is complementary to results from LEU experiments. While commercial reactors provide more intense neutrino fluxes, our analysis is completely independent on flux models related to other main fuel isotopes. In the near future, one could even imagine to use this \U~spectrum as an external constraint in analyses aiming to separate U and Pu contributions in LEU spectra \cite{DB-unfolding}.

\section{Summary}
\hspace{0.5cm} Since the emergence of the RAA, an intense experimental effort has developped around very-short-baseline reactor neutrino detectors, in order to search for active-to-sterile oscillations. A decade later, the best-fit parameters and a large portion of the allowed region in parameter space are ruled out by \STEREO, PROSPECT or DANSS. With these segmented detectors, model-independent analyses have been performed by comparing the antineutrino spectra from several baselines, and no sign of oscillations have been found.

\begin{figure} [!ht]
\centering
\centerline{\includegraphics[width=0.485\linewidth]{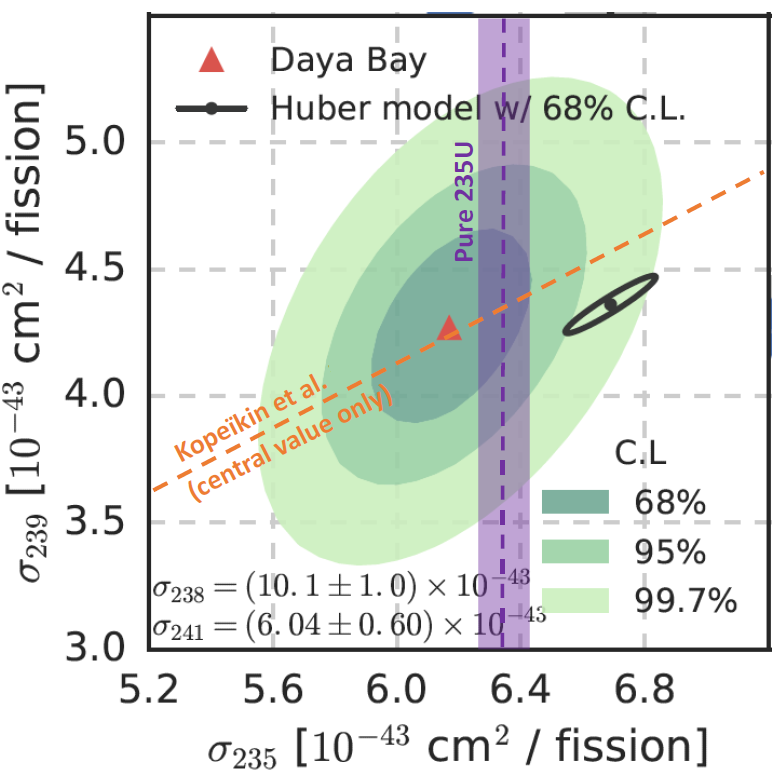}}
\caption[]{Global picture of the \U~and $^{239}$Pu cross-section per fission. The black ellipse corresponds to the HM model; the green contours are from the U/Pu separation from Daya Bay \cite{DB-U-Pu} ; the purple band reprensents the world-average for pure-\U~experiments \cite{RatePaper}; the orange dashed line corresponds to Kopeikin's new measurement of the ratio of $\beta$ spectra \U~/ $^{239}$Pu \cite{Kopeykin}.}\label{fig:kopeykin}
\end{figure}

Another explanation is then required to understand the data-to-prediction deficit of about 5\%, which first suggested the hypothesis of a sterile neutrino. The contributions of \U~and $^{239}$Pu to this deficit have been separated by the Daya Bay collaboration \cite{DB-U-Pu} and favor that the deficit is mostly carried by \U. The \U~deficit is measured at $(7.8 \pm 2.7)\%$. The measurement by \STEREO~on a pure \U~flux yields a compatible deficit of  $(5.2 \pm 2.4)\%$, with a pure-\U~world-average now at $(5.0 \pm 1.3)\%$ \cite{RatePaper}. Finally, a recent repetition by Kopeikin {\em et al.} of the measurement of \U~and $^{239}$Pu $\beta$ spectra, used as inputs for the Huber-Mueller model, indicates that the ratio of \U/$^{239}$Pu fluxes may have been overestimated by 5.4\% \cite{Kopeykin}. The global picture is illustrated in Fig.~\ref{fig:kopeykin}: it suggests that the $^{239}$Pu normalization may be correct, and the \U~normalization overestimated in the HM model by 5-6\%.

Concerning the shape anomaly, the complementarity of HEU and LEU experiments is once again an asset for the community. First revealed by reactor neutrinos experiments at commercial reactors, the "5 MeV bump" has been studied by \STEREO~and PROSPECT on pure-\U~fluxes. Combining all analyses, there is a small preference for a bump equally shared by \U~and $^{239}$Pu. As the modelling of these neutrino fluxes has proven to be a difficult task, \STEREO~and PROSPECT provide a experimental reference for the \U~spectrum. Thanks to well controlled unfolding methods, it is expressed in antineutrino energy. 

Finally, let us note that the accuracy on all these observables will increase as more data is currently taken and/or analyzed by \STEREO, PROSPECT and DANSS.


\section*{References}

\begin{thebibliography}{99}
\bibitem{RAA}G. Mention {\it et al}, \Journal{\PRD}{83}{073006}{2011}.

\bibitem{Mueller}Th.A. Mueller {\it et al}, \Journal{\PRC}{83}{054615}{2011}.

\bibitem{Huber}P. Huber, \Journal{\PRC}{84}{024617}{2011}.

\bibitem{DB} F.P. An {\it et al}, \Journal{{\em Chin. Phys.} C}{41}{013002}{2017}.
\bibitem{RENO} Z. Atif {\it et al}, arXiv:2010.14989 (2020).
\bibitem{DC} H. de Kerret {\it et al}, \Journal{\em Nature Physics}{16}{558-564}{2020}.



\bibitem{ST}N. Allemandou {\it et al}, \Journal{\JINST}{13}{P07009}{2018}.
\bibitem{PR}J. Ashenfelter {\it et al}, \Journal{\NIMA}{922}{287-309}{2019}.

\bibitem{DANSS}I. Alekseev {\it et al}, \Journal{\JINST}{11}{P11011}{2016}.

\bibitem{FIFRELIN}H. Almaz\'an {\it et al}, \Journal{{\it Eur. Phys. J.} A}{55}{183}{2019}.
\bibitem{RatePaper}H. Almaz\'an {\it et al}, \Journal{\PRL}{125}{201801}{2020}.


\bibitem{LongPaper}H. Almaz\'an {\it et al}, \Journal{\PRD}{102}{052002}{2020}.


\bibitem{PRosc}M. Andriamirado {\it et al}, \Journal{\PRD}{103}{032001}{2021}.

\bibitem{DANSSosc} M. Danilov, arXiv : 2012.10255 (2020).
\bibitem{DANSSanalysis} N. Skrobova, \Journal{\it J. Phys. Conf. Ser.}{1390}{012072}{2019}.

\bibitem{GCV} G.H. Golub, M. Heath and G Wahba, \Journal{\it Technometrics}{21:2}{215-223}{1979}.
\bibitem{ShapePaper}H. Almaz\'an {\it et al}, \Journal{\JOPG}{48}{075107}{2021}.

\bibitem{SM} M. Estienne {\it et al}, \Journal{\PRL}{123}{022502}{2019}.


\bibitem{WienerSVD} W. Tang {\it et al}, \Journal{\JINST}{12}{P10002}{2017}.
\bibitem{DB-unfolding} F.P. An {\it et al}, arXiv : 2102.04614 (2021).

\bibitem{DB-U-Pu} F.P. An {\it et al}, \Journal{\PRL}{118}{251801}{2017}.
\bibitem{Kopeykin} V. Kopeikin, M. Skorokhvatov and O. Titov, arXiv : 2103.01684 (2021).


\end{thebibliography}
\end{document}